\documentclass[aip, jmp, reprint]{revtex4-1}

\usepackage{graphicx}% Include figure files
\usepackage{natbib}
\usepackage{multirow}
\usepackage{amsmath, amssymb}
\usepackage{graphicx}
\usepackage{hyperref}
\hypersetup{
    colorlinks=true,
    linkcolor=blue,
    citecolor=blue,
    urlcolor=blue,
    pdfhighlight=/P,
    pdfauthor={Sidharath Jain},
    pdfcreator={LaTeX},
    pdftitle={Comparison of the Performance of Lenses Designed Using Field Transformation, Transformation Optics and Ray Optics},
    pdfkeywords={Flat lens}%,
}

\begin{document}

\title{Flat-lens design using Field Transformation and its comparison with those based on Transformation Optics and Ray Optics}% Force line breaks with \\

\author{Sidharath~Jain}
\email[sxj25@psu.edu]{} \affiliation{Electromagnetic Communication Laboratory, Department
of Electrical Engineering, The Pennsylvania State University,
University Park, PA-16802, USA.}

\author{Mohamed~Abdel-Mageed}
{} \affiliation{Electromagnetic Communication Laboratory, Department
of Electrical Engineering, The Pennsylvania State University,
University Park, PA-16802, USA.}
\author{Raj~Mittra}
\email[rajmittra@ieee.org]{} \affiliation{Electromagnetic Communication Laboratory, Department
of Electrical Engineering, The Pennsylvania State University,
University Park, PA-16802, USA.}

%\author{Mohamed~Abdel-Mageed}
%\author{Raj Mittra}

\begin{abstract}
This paper proposes a technique for designing flat lenses using Field Transformation (FT), as opposed to Ray Optics (RO) or Transformation Optics (TO). The lens design consists of $10$ layers of graded index dielectric in the radial direction and $5$ layers in the longitudinal direction. The central layer in the longitudinal direction primarily contributes to a bulk of the phase transformation, while  the other four layers, above and below this middle layer on either side, act as matching layers that help reduce the reflections introduced by the impedance mismatch at the interfaces of the middle layer. The paper compares the performance of the lens, so designed, with those based on the RO and TO techniques. We show that the proposed lens design using field transformation is broadband, has a better than $1$ dB higher gain compared to the RO and TO based designs over a wider frequency band, and that its scan capability is superior as well.
\end{abstract}

%\pacs{pacs numbers}

\maketitle

\textbf{Introduction:} Recently, transformation optics (TO) has been used by a number of researchers \cite{Hao2011, Smith2009} for the design of flat lenses, and it has been pointed out that such designs often call for the use of Metamaterials that may be difficult to realize in practice, especially when the required values of relative permeability and permittivity are less than unity, and/or the material properties are anisotropic. To mitigate this problem, previous designs of lenses using the TO rely on the quasi-conformal transformation optics (QCTO) approximation to reduce the anisotropy of the lens medium at the cost of making the design polarization dependent \cite{Smith2009}. However, this strategy does not obviate the need to use both the $\epsilon_r$ and $\mu_r$, even if isotropic, to provide the impedance match at the interfaces of the lens. To circumvent the practical difficulty of locating magnetic materials with the requisite values dictated by the TO, it is not uncommon to set the value of $\mu_r$ to $1$ and to vary only the $\epsilon_r$ to achieve the required refractive index  $n=\sqrt(\epsilon_r \mu_r)$, albeit at the cost of decreasing its efficiency \cite{smith2012}. A plano-concave lens has recently been designed by using Metamaterials, to realize a gain above $13$ dB in the frequency band ranging from $10$ to $12$ GHz \cite{MIT_2012}. However, such a lens has a narrow bandwidth, which is typically the case of metamaterial-based designs, especially those with resonant inclusions. Of course, the conventional flat lens designs, that are based on the Ray Optics approach, do not suffer from the same drawbacks as encountered in the TO designs; however, they do not offer the same level of flexibility in controlling the amplitudes and phases of the fields traversing through the lens from the input to the output port, as does the Field Transformation (FT) method that we propose herein.

%The TO algorithm is based on geometry transformation, which preserves the field variation when we map the physical system to the virtual one, or vice versa. It does provide us the information on the material parameters ($\epsilon_r$,$\mu_r$) in the physical system from the knowledge of those in the virtual system. The caveat is, though, that sometimes these values, dictated by the TO, can be difficult to realize in practice because they maybe either less than 1; or much greater than 1. The quasi-conformal transformation optics (QCTO) approximation has been proposed as a way to reduce the anisotropy of the medium and also enables one to assume that $\mu_r = 1$ albeit at a cost.
\begin{figure}[h]
\begin{center}
\noindent
  \includegraphics[width=1.5in]{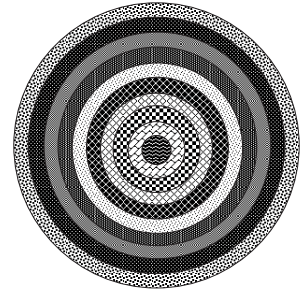}
  \caption{Multilayered Lens Cross-section}\label{Fig1}\vspace{-.2in}
\end{center}
\end{figure}
\begin{figure}[h]
\begin{center}
\noindent
  \includegraphics[width=2.5in,height=.7in]{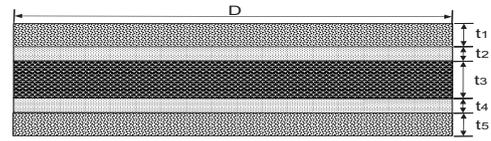}
  \caption{Multilayered Lens Side View}\label{Fig2}\vspace{-.2in}
\end{center}
\end{figure}
%\begin{figure}[h]
%\begin{center}
%\noindent
%  \includegraphics[width=3.0in]{grin.eps}
%  \caption{GRIN Lens}\label{Fig3}
%\end{center}
%\end{figure}
%\begin{figure}[h]
%\begin{center}
%\noindent
%  \includegraphics[width=3.0in]{zoneplate.eps}
%  \caption{GRIN Zone Plate Lens}\label{Fig4}
%\end{center}
%\end{figure}
%\begin{figure}[h]
%\begin{center}
%\noindent
%  \includegraphics[width=3.0in]{luneburg.eps}
%  \caption{Cross Section of a Spherical Luneburg Lens}\label{Fig5}
%\end{center}
%\end{figure}
\begin{figure}[!ht]
\begin{center}
\noindent
  \includegraphics[width=1.7in]{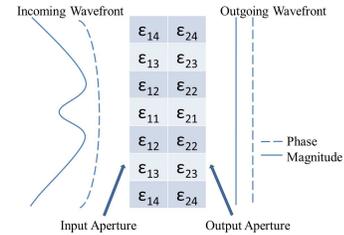}
  \caption{Field Transformation Principle}\label{Fig6}
\end{center}
\end{figure}

\begin{figure}[!ht]
\begin{center}
\noindent
  \includegraphics[width=3in]{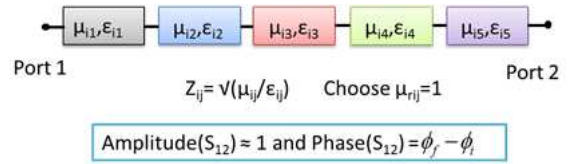}
  \caption{Optimal Filter Design}\label{Fig10}\vspace{-.3in}
\end{center}
\end{figure}

%\begin{figure}[!ht]
%\begin{center}
%\noindent
%  \includegraphics[width=1.5in]{design.eps}
%  \caption{Design principle}\label{Fig7}\vspace{-.3in}
%\end{center}
%\end{figure}

\textbf{Lens Design:} We begin with the Graded Index (GRIN) approach to designing a flat lens by using the concept of field transformation.  The basic concept upon which this approach is based is very similar to that proposed by Luneburg for the design of the spherical lenses \cite{luneburg}. Taking a cue from Luneburg, we simply specify the desired field distribution in the output port (exit aperture) and determine the medium parameters of the intervening medium such that the given field distribution in the input port (input aperture) is transformed to the desired field distribution in the exit plane, as shown in Fig.\ref{Fig6}. We can accomplish this task, in many cases, by tracing rays through the inhomogeneous medium that we are trying to synthesize. The design parameters of the lens are its center frequency ($f$), focal length ($F$), thickness ($t$) and gain. Next, we choose the diameter (D) of the lens depending on the gain requirements, and the level of discretization of the material parameters along the radial direction, which determines the number of rings. For the flat lens design, shown in Figs. \ref{Fig1} and \ref{Fig2}, we assign $f = 30$ GHz, $\Sigma{t_i}=t=10$ mm, $D=63.5$ mm and $10$ discrete rings each with a width of $3.175$ mm, in order to facilitate later comparison of its performance with that of the TO design \cite{Hao2011}.
\begin{figure}[h]
\begin{center}
\noindent
  \includegraphics[width=1.5in]{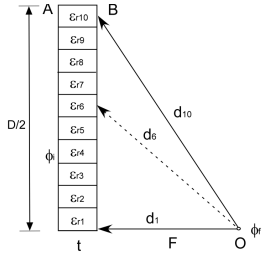}
  \caption{Design principle}\label{Fig7}\vspace{-.3in}
\end{center}
\end{figure}

\begin{figure}[!ht]
\begin{center}
\noindent
  \includegraphics[width=6cm]{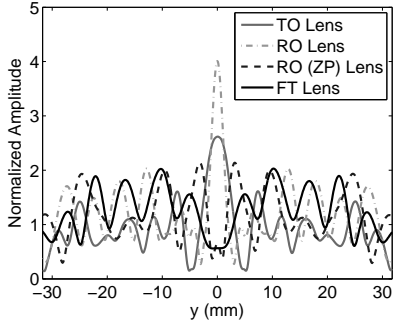}
  \caption{Comparison of amplitude distribution at the aperture}\label{Fig3A}\vspace{-.3in}
\end{center}
\end{figure}

\begin{figure}[!ht]
\begin{center}
\noindent
  \includegraphics[width=6cm]{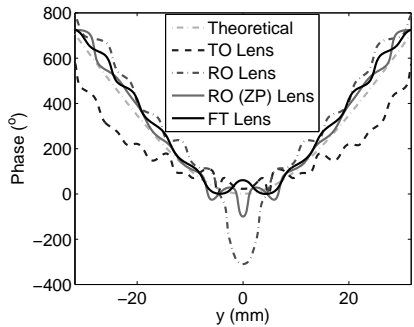}
  \caption{Comparison of phase distribution at the aperture}\label{Fig3B}\vspace{-.3in}
\end{center}
\end{figure}

\begin{figure}[!ht]
\begin{center}
\noindent
  \includegraphics[width=6cm]{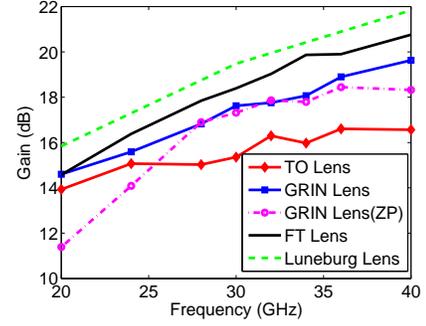}
  \caption{Comparison of Gain (in dB) as a function of frequency}\label{Fig8}\vspace{-.3in}
\end{center}
\end{figure}

\begin{table}[hb]
\begin{center}
\renewcommand{\arraystretch}{1.4}
\caption{Material Parameters of the FT Lens in Fig.\ref{Fig2}} \label{Table1}
\begin{tabular}{c|c c c c c c c c c c c}
 \hline\hline
 $\bf{\epsilon_{rij}}$&	\bf{1}&	\bf{2}&	\bf{3}	&\bf{4}	&\bf{5}	&\bf{6}	&\bf{7}	&\bf{8}	&\bf{9}	&\bf{10}	&\bf{$t_i$}\\
   \hline
 \bf{1}	&1.6&	1.55	&1.55&	1.58&	1.58&	1.58&	1.5	&1.5	&1.5	&1.45	& 1.94\\

 \bf{2}	&9.5	&8.82	&7.92	&7.90	&7.89	&7.88	&6.52	&3.98	&2.20	&1.60	&.79\\

\bf{3}	&25.5	&24.5	&22.3	&18.5	&14.55	&10.5	&7.65	&5.5	&3.5	&1.65	&4.01\\

\bf{4}	&9.5	&8.82	&7.92	&7.90	&7.89	&7.88	&6.52	&3.98	&2.20	&1.60	& 0.79\\

\bf{5}	&1.6	&1.55	&1.55	&1.58	&1.58	&1.58	&1.5	&1.5	&1.5	&1.45	&1.94\\
 \hline\hline
\end{tabular}
\end{center}\vspace{-.2in}
\end{table}

\begin{table}[ht]
\begin{center}
\renewcommand{\arraystretch}{1.4}
\caption{Material Parameters of the RO Lens} \label{Table2}
\begin{tabular}{c c c c c c c c c c}
 \hline\hline
 $\bf{\epsilon_{r1}}$&	 $\bf{\epsilon_{r2}}$&$\bf{\epsilon_{r3}}$&$\bf{\epsilon_{r4}}$&$\bf{\epsilon_{r5}}$&$\bf{\epsilon_{r6}}$&$\bf{\epsilon_{r7}}$&$\bf{\epsilon_{r8}}$&$\bf{\epsilon_{r9}}$&$\bf{\epsilon_{r10}}$\\
   \hline
11.2&	10.60&	9.80&	8.70&	7.45&	6.15&	4.88&	3.70&	2.64&	1.73\\
 \hline\hline
\end{tabular}
\end{center}\vspace{-.2in}
\end{table}

%\begin{figure}[hb]
%\centering
%\subfigure[$20$ GHz] % caption for subfigure b
%{
%   \label{fig9:a}
%   \includegraphics[width=6cm]{gainvsangle_20.eps}
%} \hspace{-0.5cm}
%\subfigure[$30$ GHz]  % caption for subfigure a
%{
%   \label{fig9:b}
%   \includegraphics[width=6cm]{gainvsangle_30.eps}
%}\hspace{-0.5cm}
%\subfigure[$40$ GHz] % caption for subfigure b
%{
%   \label{fig9:c}
%   \includegraphics[width=6cm]{gainvsangle_40.eps}
%} \caption{Comparison of Gain (in dB) as a function of scan angle for three different frequencies.}\vspace{-.3in}
%\label{fig9} % caption for the whole figure
%\end{figure}
\begin{figure}[hb]
\begin{center}
\noindent
   \includegraphics[width=6cm]{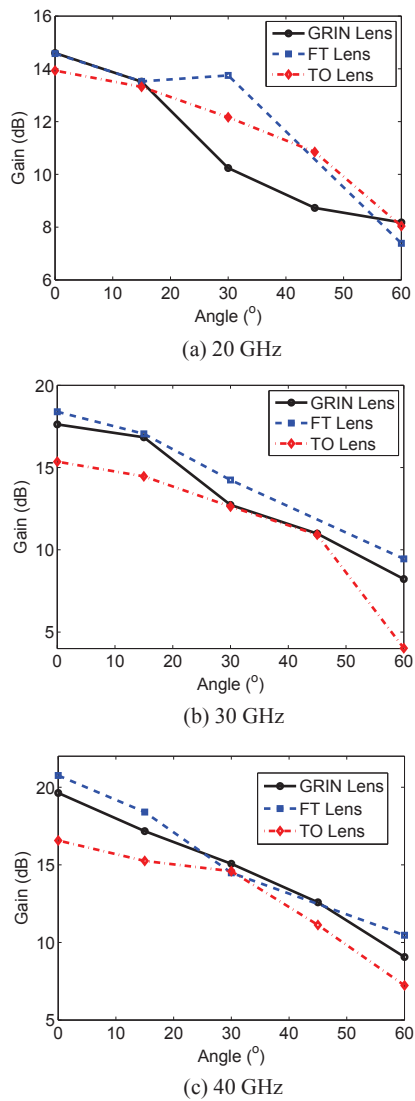}
 \caption{Comparison of Gain (in dB) as a function of scan angle for three different frequencies.} \label{fig9} \vspace{-.3in}
% caption for the whole figure
\end{center}
\end{figure}

\begin{table}[ht]
\begin{center}
\renewcommand{\arraystretch}{1.4}
\caption{Material Parameters of the RO Zone Plate Lens} \label{Table3}
\begin{tabular}{c c c c c c c c c c}
 \hline\hline
 $\bf{\epsilon_{r1}}$&	 $\bf{\epsilon_{r2}}$&$\bf{\epsilon_{r3}}$&$\bf{\epsilon_{r4}}$&$\bf{\epsilon_{r5}}$&$\bf{\epsilon_{r6}}$&$\bf{\epsilon_{r7}}$&$\bf{\epsilon_{r8}}$&$\bf{\epsilon_{r9}}$&$\bf{\epsilon_{r10}}$\\
  \hline
 4.96&	4.62&	4.08&	3.38&	2.62&	1.88&	1.21&	3.70&	2.64&	1.73\\
 \hline\hline
\end{tabular}
\end{center}\vspace{-.3in}
\end{table}

\begin{table}[ht]
\begin{center}
\renewcommand{\arraystretch}{1.4}
\caption{Material Parameters of the Spherical Luneburg Lens} \label{Table4}
\begin{tabular}{c c c c c c c c c c c}
 \hline\hline
 $\bf{\epsilon_{r1}}$&	 $\bf{\epsilon_{r2}}$&$\bf{\epsilon_{r3}}$&$\bf{\epsilon_{r4}}$&$\bf{\epsilon_{r5}}$&$\bf{\epsilon_{r6}}$&$\bf{\epsilon_{r7}}$&$\bf{\epsilon_{r8}}$&$\bf{\epsilon_{r9}}$&$\bf{\epsilon_{r10}}$&$\bf{\epsilon_{r11}}$\\
   \hline
 2.0&	1.96&	1.92&	1.86&	1.78&	1.68&	1.56&	1.43&	1.28&	1.11&	1.05\\
 \hline\hline
\end{tabular}
\end{center}\vspace{-.3in}
\end{table}

%
%\begin{table*}[ht]
%\begin{center}
%\renewcommand{\arraystretch}{1.4}
%\caption{Comparison of the electric field enhancement in the focal region as a function of frequency for a normally incident plane wave} \label{Table5}
%\begin{tabular}{|c|c|c|c|c|c|}
% \hline
% Frequency
%   (in GHz)&	TO Lens&	RO Lens&	RO (ZP) Lens&	FT Lens &Luneburg Lens\\
%\hline
%20&	4.974&	5.37	&3.71	&5.358	&6.196\\
%\hline
%24&	5.668&	6.018	&5.064	&6.593&\\
%\hline	
%28&	5.641&	6.936	&6.998	&7.803&\\
%\hline	
%30&	5.86&	7.606	&7.342	&8.309	&9.418\\
%\hline
%32&	6.53&	7.725	&7.821	&8.944&\\
%\hline	
%34&	6.291&	8.006	&7.756	&9.85&\\
%\hline	
%36&	6.764&	8.81	&8.362	&9.884&\\
%\hline	
%40&	6.733&	9.584	&8.241	&10.91	&12.35\\
% \hline
%\end{tabular}
%\end{center}
%\end{table*}

Once the design parameters have been selected, the phase required at the output face of the lens is calculated using the condition that a plane wave incident upon the input face interferes constructively at the focal point of the lens. Then a cascade of transmission lines with different characteristic impedances, whose aggregate thickness is equal to $t$ are used to realize the required phase and to maximize the transmission coefficient over a wide frequency band, as shown in Fig.\ref{Fig10}. Also, the layers are chosen to be symmetric about the central one, so that the performance of the lens would be similar for waves incident on either of its flat faces. We mention here that it is neither convenient nor straightforward to control the impedance match aspect of the design in the context of the TO, especially because we restrict the permeability $\mu$ to be equal to $\mu_0$.

Let $\phi_i$ be the phase of the plane wave incident from the left on the face A of the half-lens in Fig.\ref{Fig7} which we are considering by taking advantage of the symmetry of the lens. Then, for the rays from different portions of the lens to constructively interfere at the focal point $O$, the electrical length of all the waves traveling from face A of the lens to the point $O$ should either be the same or should differ by an integral multiple of the wavelength, i.e.,

\begin{equation}
\label{eq:eq1}
\phi_f-\phi_i = k_0(t\sqrt{\epsilon_{rij}}-n\lambda_0+d_j)
\end{equation}

where $k_0$ and $\lambda_0$ are the free-space wavenumber and wavelength, respectively; $n$ is an integer; $d_j$ is the free-space path difference between the center of $j^{th}$ ring of the lens and the focal point O; and $\epsilon_{rij}$ is the dielectric constant of the $j^{th}$ ring located in the $i^{th}$ layer of the lens. The rings ($i$) in Fig.\ref{Fig1} are numbered from $1$ to $10$ from the central to the outermost one. The layers ($j$) in Fig.\ref{Fig2} are numbered from $1$ to $5$ from the top to the bottom.

Our goal is to maximize the performance of the lens and to achieve this goal, we attempt to realize the desired phases on the face B of the lens, while simultaneously maximizing the transmission coefficient over a broad frequency band. We can solve this problem by using a transmission line model in a circuit simulator to design an optimum multilayer filter, comprising of $5$ layers, which achieves the desired phase at the center frequency and a transmission coefficient as close to $1$ as possible over a broad frequency band for each of the $10$ discrete rings shown in Fig.\ref{Fig1}. The middle layers perform a majority of the phase transformation, while the other two layers on either side act as matching layers in order to maximize the transmission when the wave is incident from either side. This enables us to design a lens for a specified gain and bandwidth. The material parameters obtained by following the above procedure are given in Table \ref{Table1}. The materials required to fabricate the lens can be synthesized by using a number of techniques for manipulating the dielectric properties of materials that have been developed recently \cite{Vardax,eff_medium,3Dprinting,3Dprinting1,Zhang}. Before closing we mention that the FT approach can also be used to further control the amplitude distribution in the exit aperture plane, with a view to maximizing the aperture efficiency of the lens, a feature that is not readily available in the TO or RO designs.

When the required focal length of the lens is small, then the phase of the wave on the face B of the lens would have a range that exceeds $360^o$. In this case, it is possible to subtract an integral multiple of $360^o$ from the desired phase so as to reduce the range of dielectric constant needed to realize the lens. It is evident that this is equivalent to reducing the net electrical length by an integral multiple of the wavelength. In this work, we refer to the lenses designed in this manner as Zone Plate (ZP) lenses. Such zone plating of a lens helps avoid the need to use very high values of $\epsilon_{\text{r}}$, albeit at the cost of reducing the bandwidth of the lens.

We have also designed a $10$-ring RO lens and a $10$-ring RO (ZP) lens, with material parameters listed in Table \ref{Table2} and \ref{Table3} in order to show the advantages of the proposed FT technique over other flat lens design techniques that have appeared in the literature. A $11$-layer spherical Luneburg lens whose material parameters are listed in Table \ref{Table4} was also designed. Inorder to compare our results with the lenses based on the TO paradigm, we also designed an $11$-layer lens using the material parameters given in \cite{Hao2011} which are obtained by transforming a hyperbolic lens into a flat lens using the TO algorithm. To make a fair comparison the diameters and the widths of all these lenses were chosen to be identical.

\textbf{Results:} Fig.\ref{Fig3A} shows a comparison of the amplitude distribution at the aperture of the four different lens designs. The FT lens has a close to flat amplitude distribution at the aperture but the RO lens, the RO (ZP) lens and the TO lens suffer from a mismatch and hence the amplitude of the wave at the exit aperture is uneven. The amplitude of the TO lens is even lower than that of the RO-based designs, possibly because of higher mismatch. Fig.\ref{Fig3B} shows the phase distribution along a diameter of the exit aperture. The phase distribution of the FT lens is closest to the ideal phase distribution. The phase of the RO design is also close to the ideal phase, but the phase of the TO design using the material parameters in \cite{Hao2011} deviates from the ideal as one moves away from the center of the lens, which can be attributed to the isotropic, nonmagnetic and non-metallic medium approximations used to fabricate practical devices based on the TO approach. The phase distribution for the RO (ZP) lens distorts as we move away from the center frequency, as expected.  This, in turn, makes this lens relatively narrowband.

Fig.\ref{Fig8} shows the gain (in dB), defined as $20$ log$_{10}(|E_f|/|E_i|)$, where $E_f$ is electric field at the focal point $O$ and $E_i$ is the electric field of the incident plane wave. The gain of the proposed FT lens has been compared with the TO lens \cite{Hao2011}, RO lens, RO (ZP) Lens and the spherical Luneburg lens. The Luneburg lens has the best performance as expected but the multilayered lens has more than $1$ dB higher gain compared to the remaining three designs over the frequency band from $20$ to $40$ GHz.  The gain values of the three competent flat lens designs, namely, the FT lens, the RO lens and the TO lens, were also compared for three different frequencies, namely. $20$ GHz, $30$ GHz and $40$ GHz, for the scan angle varying from $0^o$ to $60^o$. The results for these three frequencies  are shown in Fig. \ref{fig9}. In addition, we observe that the FT lens performs equally well when the wave is incident from either side, while the performance of the TO lens deteriorates when the wave is incident on the face that has a higher dielectric constant, perhaps because of higher impedance mismatch.

\textbf{Conclusion:} In this paper the field transformation (FT) technique has been proposed and has been used to design a five layered flat lens. The performance of the FT lens has been compared with that of lenses designed by using ray optics (GRIN and GRIN zone plate) and transformation optics (TO) The FT lens  has been found to have at least $1$dB higher field enhancement, over the RO and TO designs, in the focal region over a broad bandwidth. Also, the designed FT lens has a higher field enhancement in the focal region, and over a larger range of scan angles, when compared with competing designs. Finally, it has been pointed out that the Field Transformation approach does not suffer from the realizability issues encountered in TO-based designs, that are based on geometry transformation instead.  In fact, an important advantage of the FT approach, besides its simplicity, is that we can utilize realistic dielectric and magnetic materials of our choice that are conveniently realizable by using available materials, including non-magnetic materials, and work only with isotropic materials if that is our preference. Furthermore, we have the flexibility to not only optimize the desired phase, but also the transmission through the system, by introducing suitable matching layers at the input and output planes, for instance, to ensure maximum transmission accompanied by low reflection losses. We can further embellish this technique to control both the amplitude and the phase distributions in an aperture, which is difficult to do using other approaches.

\bibliographystyle{aipnum4-1}

%\bibliography{multilayered_complete_apl}
%\nocite{*}

\end{document}